\documentclass[conference]{IEEEtran}
\IEEEoverridecommandlockouts

\usepackage{graphicx}
\usepackage{cite}
\usepackage{amsmath,amssymb,amsfonts}
\usepackage[mathscr]{euscript}
\usepackage{mathtools, nccmath}
\usepackage{float}
\usepackage{algorithmic}
\usepackage{graphicx}
\usepackage{textcomp}
\usepackage{xcolor}
\usepackage{bm}\newcommand{\minus}{\scalebox{0.75}[1.0]{$-$}}

\makeatletter
\newenvironment{taggedsubequations}[1]
 {%
  \addtocounter{equation}{-1}%
  \begin{subequations}%
  \def\@currentlabel{#1}%
  \renewcommand{\theequation}{#1.\arabic{equation}}%
 }
 {\end{subequations}}
\makeatother 

\usepackage[linesnumbered,ruled,vlined]{algorithm2e}
\let\oldnl\nl
\newcommand{\nonl}{\renewcommand{\nl}{\let\nl\oldnl}}
\makeatletter
\SetKwBlock{Repeat}{repeat}{}

\def\BibTeX{{\rm B\kern-.05em{\sc i\kern-.025em b}\kern-.08em
    T\kern-.1667em\lower.7ex\hbox{E}\kern-.125emX}}
\makeatletter
\let\old@ps@headings\ps@headings
\let\old@ps@IEEEtitlepagestyle\ps@IEEEtitlepagestyle
\def\confheader#1{%

	\def\ps@IEEEtitlepagestyle{%
		\old@ps@IEEEtitlepagestyle%
		\def\@oddhead{\parbox[b]{9cm}{\raggedright \@IEEEheaderstyle #1}}%
		\def\@evenhead{\strut\hfill#1\hfill\strut}%
	}%
	\ps@headings%
}
\makeatother

\confheader{%
	\textbf{\textit{2023 IEEE International Conference on Energy Technologies for Future Grids \\                                                                                                     
			Wollongong, Australia, December 3-6, 2023}}
}

\begin{document}

\title{Network-aware EV charging and discharging in unbalanced distribution grids: A distributed, robust approach against communication failures
\thanks{This work was supported by the Australian Research Council under Grant DP220101035.}
}

\author{\IEEEauthorblockN{Nanduni Nimalsiri}
\IEEEauthorblockA{
\textit{The Australian National University}\\
Canberra, Australia \\
nanduni.nimalsiri@anu.edu.au}
\and
\IEEEauthorblockN{Elizabeth Ratnam}
\IEEEauthorblockA{\textit{The Australian National University}\\
Canberra, Australia \\
elizabeth.ratnam@anu.edu.au}
}

\IEEEoverridecommandlockouts
\IEEEpubid{\makebox[\columnwidth]{978-1-6654-7164-0/23/\$31.00~\copyright2023 IEEE  \hfill} \hspace{\columnsep}\makebox[\columnwidth]{ }}
\maketitle
\IEEEpubidadjcol

\begin{abstract}
This paper proposes a distributed optimization-based algorithm for electric vehicle (EV) charging and discharging, incorporating EV customer economics and distribution network constraints enforced on an unbalanced distribution grid. Building on a consensus-based alternating direction method of multipliers (ADMM), the algorithm is designed such that EVs coordinate by means of exchanging limited information with their neighboring EVs in a connected communication network. Specifically, an iterative routine is executed, whereby EVs cooperatively determine their charge-discharge profiles that maintain the distribution grid voltages and transformer core temperatures within safe operating limits. Importantly, the algorithm is robust against communication failures potentially arising in real-world implementations. Numerical simulations are conducted to verify the efficacy of the proposed EV charging algorithm in terms of network-aware operation and communication-failure tolerant operation.

\end{abstract}

\begin{IEEEkeywords}
ADMM, Communication failures, Distributed control, Electric vehicle charging, Network-aware.
\end{IEEEkeywords}

\section{Introduction}

As the electrification of transportation systems continues to displace fossil fuel-based vehicles with electric vehicles (EVs), there is an increased urgent need for controlling and coordinating EV charging (and discharging) to minimize electric grid congestion \cite{nimalsiri2019survey}. Specifically, grid-connected EVs must be strategically coordinated via distribution \textit{network-aware} approaches to mitigate undesirable voltage excursions and thermal overloads arising from their charging operations.

Numerous optimization-based coordination approaches have been proposed for network-aware EV charging. However, only a limited number of works (e.g., \cite{zhang2016scalable, nimalsiri2021distributed, bhardwaj2022communication}) have focused on \textit{unbalanced} distribution grids. The recent literature on network-aware EV charging has considered approaches with \textit{centralized control} (all EV charge profiles are determined by a single entity as in \cite{nimalsiri2021coordinated}) and \textit{distributed control} (EV charge profiles are determined by multiple entities via shared and parallelized computations as in \cite{nimalsiri2021distributed, bhardwaj2022communication, zhang2016scalable, liu2017decentralized, rahman2021continuous, zhou2020voltage}) --- with the latter becoming vital to enable EV proliferation. The majority of existing distributed EV charging algorithms as in \cite{liu2017decentralized, zhang2016scalable, zhou2020voltage, rahman2021continuous} have employed a star communication network topology in which all information is transmitted to EVs via an aggregator, and as such, they are susceptible to communication-based single point of failures. Alternatively,  the \textit{fully distributed} EV charging algorithms in \cite{nimalsiri2021coordinated, wang2016fully} have eliminated the aggregator and instead have allowed EVs to communicate amongst themselves. Such implementations however result in an excessive communication overhead when each EV has to communicate with all the other EVs in the communication network. Therefore, approaches involving peer-to-peer communication with a subset of EVs (in the vicinity) as in \cite{nimalsiri2021distributed, bhardwaj2022communication} are comparatively efficient in terms of both computation and communication. 

The \textit{Alternating Direction Method of Multipliers} (ADMM) \cite{boyd2011distributed} has been widely applied in various fields to develop computationally efficient distributed algorithms. In \cite{nimalsiri2021distributed}, the authors have proposed an ADMM-based EV charging algorithm to regulate the supply voltages to remain within operational limits. A practical limitation that exists in the algorithm in \cite{nimalsiri2021distributed} is that it cannot handle \textit{communication failures} (e.g., disconnection of agents or packet losses in message exchange) that are more frequent in real-world implementations. Consequently, we are motivated to address the robustness in case of communication failures potentially arising when coordinating EVs via distributed communication.

In this paper, we propose a fully-distributed algorithm that coordinates EVs to alleviate the impacts of EV charging and discharging on an unbalanced distribution grid, including under- and over-voltage conditions and thermal overload conditions. The underlying EV charging optimization problem is formulated to minimize the operational costs accrued to EV customers (i.e, cost of electricity for EV charging and discharging on a Time-of-Use (ToU) net-metering tariff and cost of battery degradation due to frequent charging and discharging), while maintaining the grid nodal voltages and transformer core temperatures within the prescribed quasi-steady-state voltage and thermal limits. The customer-specified charging requirements and battery constraints, including limitations in battery charge-discharge rates and state-of-charge thresholds, are also included in the optimization problem. The proposed distributed algorithm underpins a \textit{consensus-based ADMM} approach with EVs as computing \textit{agents} executing an \textit{iterative routine}. To develop the algorithm, we reformulate the underlying optimization problem into a decomposable form that is compliant with ADMM and then develop an iterative consensus-based approach to solve individual EV subproblems dealing with coupling network constraints. Specifically, the proposed algorithm is designed such that EVs communicate in a peer-to-peer (connected) communication network and determine their charge-discharge schedules locally without requiring any intermediary aggregators. 

The contribution of this paper is two-fold: (1) This is the first paper we are aware of that studies distributed EV charging control in an unbalanced distribution grid taking into account both nodal voltage constraints and transformer thermal constraints. (2) We facilitate operation in the presence of communication failures potentially arising in a nonideal communication network where the EV charging points can be active and inactive randomly and the communications among EVs can fail probabilistically. The algorithm additionally has the following features: it (i) requires only limited information exchange between EVs, implying lower communication overhead; (ii) improves EV user privacy by not transmitting any EV-specific private information; (iii) does not need synchronization between EVs when executing the algorithm.

\section*{Notation}

$\mathbb{R}^m$ (and $\mathbb{C}^m$) denote $m$-dimensional vectors of real  (and complex) numbers. $\mathbb{R}^{m\times n}$ (and $\mathbb{C}^{m\times n}$) denote $m$-by-$n$  matrices of real (and complex) numbers. $\mathbf{0}$ is the all-zeros column vector, $\mathbf{1}$ is the all-ones column vector, and $\mathbf{I}$ is the identity matrix, where the context will make clear the dimensions intended. For $\bm{A}\in\mathbb{R}^{n\times m}$ (or $\mathbb{C}^{n\times m}$), $[\bm{A}]_{i,j}$ denotes the entry in row $i$ and column $j$, $\bm{A}^\top\in\mathbb{R}^{m\times n}$ (or $\mathbb{C}^{m\times n}$) denotes its transpose. $\mathbf{T}$ denotes the square matrix satisfying $[\mathbf{T}]_{i,j}=1$ for $i\geq j$ and $[\mathbf{T}]_{i,j}=0$ elsewhere. Given matrices $\bm{A}\in\mathbb{R}^{m\times n}$ and $\bm{B}\in\mathbb{R}^{p\times q}$, $\bm{C}\coloneqq \bm{A}\oplus \bm{B}=\mathrm{diag}(\bm{A},\bm{B})\in\mathbb{R}^{(m+p)\times(n+q)}$.

\section{Problem Formulation}

\subsection{Residential Energy System}

\begin{figure}[t]
    \centering
    \includegraphics[width=5cm]{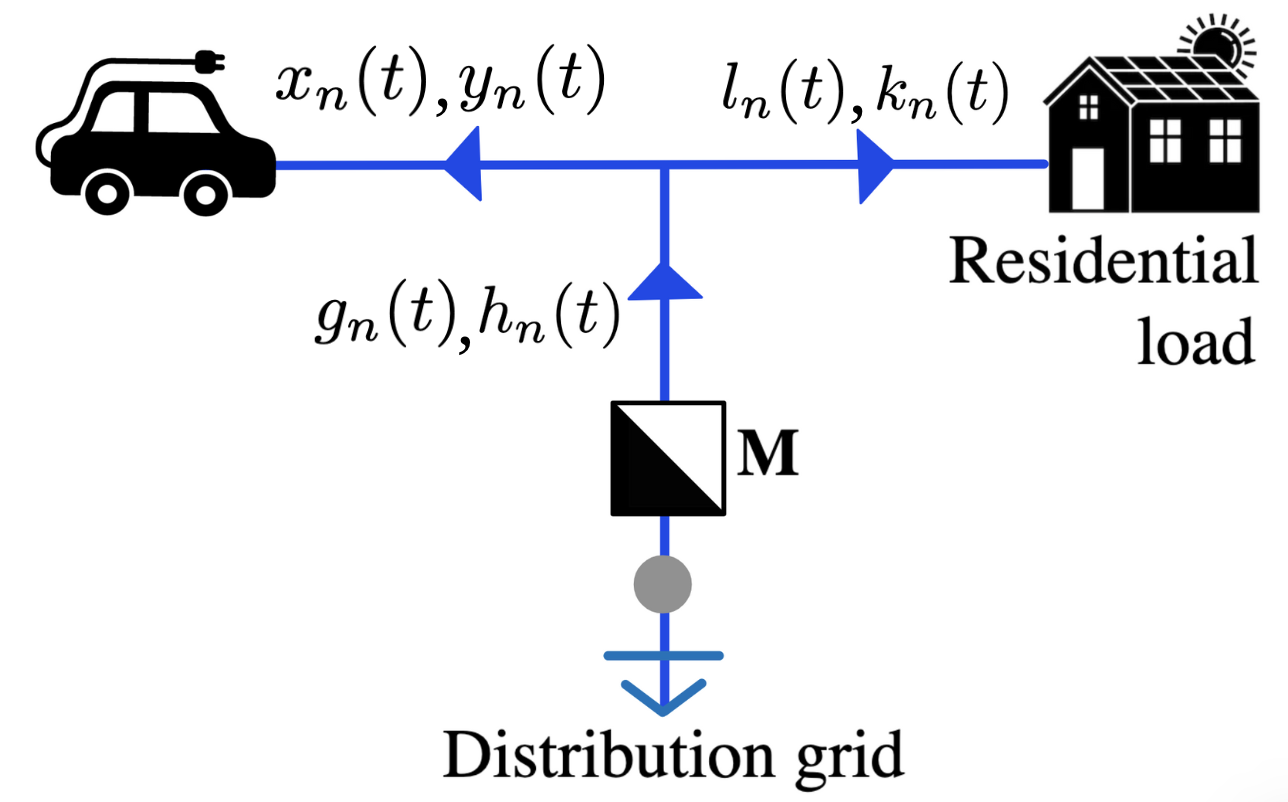}
    \caption{A Residential Energy System for customer $n \in\mathcal{N}$. Arrows indicate the direction of positive power flow. \textbf{M} is a bi-directional meter.}
    \label{fig-1}
\end{figure}

Fig.~\ref{fig-1} illustrates the topology of a grid-connected Residential Energy System (RES) for a single customer, consisting of an EV and a residential load situated behind the Point of Common Coupling (PCC). We consider a set of $N$ customers defined by $\mathcal{N}:=\{1,\dots,n,\dots,N\}$, with customer $n$ charging or discharging $\text{EV}_n$ over a discrete planning time-horizon $[0,T\Delta]$ that consists of a set of time-steps of length $\Delta$ [hours], indexed by $\mathcal{T} \coloneqq \{1, \dots, t, \dots, T\}$. Let $x_{n}(t)$ [kW] (and $y_{n}(t)$ [kVAR]) denote the real power (and reactive power) to $\text{EV}_n$ at time $t\Delta$. That is, $x_n(t)> 0$ when charging, $x_n(t)< 0$ when discharging, and $x_n(t)= 0$ when remaining idle. Let $l_n(t)$ (and $k_n(t)$) denote the real (and reactive) residential power consumption in excess or deficit of photovoltaic generation (if rooftop solar exists at home) at time $t\Delta$. The measured real (and reactive) power through meter $\textbf{M}$ at time $t\Delta$ is denoted by $g_n(t)$ (and $h_n(t)$). 

For each $\text{EV}_n$,  $\sigma_n(t)$ is the state of charge (SoC) at time $t\Delta$ (as a fraction), and $\bm{\sigma}_n \coloneqq [\sigma_n(1),\dots,\sigma_n(t),\dots,\sigma_n(T)]^\top\in\mathbb{R}^T$ is the \textit{SoC profile}. Moreover, $\bm{x}_n \coloneqq [x_n(1),\dots,x_n(t),\dots,x_n(T)]^\top\in\mathbb{R}^T$ denotes the \textit{EV charge-discharge profile}. The set of design parameters that corresponds to $\text{EV}_n$ includes the arrival time index $a_{n}\in\mathcal{T}$, departure time index $d_{n}\in\mathcal{T}$, battery capacity $c_{n}$ [kWh], initial SoC $\hat{\sigma}_n$ (such that $\sigma_n(a_n) = \hat{\sigma}_n$), target SoC at the departure $\sigma_n^*$, energy conversion efficiency of (dis)charging $\mu_n$ (\%), minimum SoC $\underline{\sigma}_n$, maximum SoC $\overline{\sigma}_n$, maximum charge rate $\overline{x}_n$ and maximum discharge rate $\underline{x}_n$. Following \cite{nimalsiri2021distributed}, $\sigma_n(t)$ is computed by $\sigma_n(t) \coloneqq \sigma_n(t-1) +  x_n(t)\mu_n\frac{1}{c_n}\Delta$. 

Next, we introduce the battery constraints for $\text{EV}_n$. To prevent over-charging (and discharging) of the battery, the SoC profile $\bm{\sigma}_n$ is constrained by $\underline{\sigma}_n\mathbf{1} \leq \bm{\sigma}_n \leq \overline{\sigma}_n\mathbf{1}$, which can be equivalently expressed by
\begin{equation}\label{cons-2}
  \underline{c}_n \mathbf{1} \leq \mathbf{T}\bm{x}_n \leq \overline{c}_n \mathbf{1},
\end{equation}
where $\underline{c}_n\coloneqq c_n(\underline{\sigma}_n-\hat{\sigma}_n)/(\mu_n\Delta)$ and $\overline{c}_n \coloneqq c_n(\overline{\sigma}_n-\hat{\sigma}_n)/(\mu_n\Delta)$.  The charging and discharging rates of the battery is limited by
\begin{equation}\label{cons-3}
   \underline{x}_n \mathbf{1} \leq \bm{x}_n \leq \overline{x}_n \mathbf{1}.
\end{equation}

The requirement to satisfy the customer's charging demand $e_n \coloneqq (\sigma_n^* - \hat{\sigma}_n)c_n$ before the EV's next departure is enforced by the constraint $\sigma_n(T) = \sigma_n^*$, which implies that
\begin{equation}\label{cons-4}
  \mathbf{1}^\top \bm{x}_n = e_n/\Delta.
\end{equation}

$\text{EV}_n$ is allowed to charge or discharge only when it is available (grid-connected) as defined by the availability matrix $\bm{\mathcal{L}}_n\in\mathbb{R}^{T\times T}$, in which $[\bm{\mathcal{L}}_n]_{i,j}=1$  if $i = j$ and $a_n < i \leq d_n$ and $[\bm{\mathcal{L}}_n]_{i,i}=0$ otherwise. Accordingly, we have
\begin{equation}\label{cons-5}
    (\mathbf{I} - \bm{\mathcal{L}}_n)\  \bm{x}_n = \mathbf{0}.
\end{equation}

Consequently, the battery constraints in \eqref{cons-2}-\eqref{cons-5} can be combined to obtain the following feasible set for $\bm{x}_n$.
\begin{equation}\label{aggregate_battery_cons}
      \Psi_n \coloneqq \{ \bm{x}_n | \ \bm{A}_n \bm{x}_n \geq \bm{b}_n, \  \bm{\bar{A}}_n \bm{x}_n = \bm{\bar{b}}_n\},
\end{equation}
where $\bm{A}_n \coloneqq [\mathbf{I} \ \   \minus\mathbf{I} \ \   \mathbf{T}^\top \ \  \minus\mathbf{T}^\top]^\top\in \mathbb{R}^{4T \times T}, \bm{\bar{A}}_n\coloneqq [\mathbf{1} \   (\mathbf{I} - \bm{\mathcal{L}}_n)]^\top\in \mathbb{R}^{(T+1)\times T}, \bm{b}_n  \coloneqq [\underline{x}_n\mathbf{1}^\top \ \     
    \minus\overline{x}_n\mathbf{1}^\top  \ \ 
    \underline{c}_n\mathbf{1}^\top \ \ 
    \minus\overline{c}_n\mathbf{1}^\top]
^\top\in \mathbb{R}^{4T}, \bm{\bar{b}}_n \coloneqq [e_n/\Delta  \ \ \mathbf{0}^\top]^\top \in \mathbb{R}^{T+1}$. Here  $\mathbf{I}\in\mathbb{R}^{T\times T}$, $\mathbf{T}\in\mathbb{R}^{T\times T}$, $\mathbf{1}\in\mathbb{R}^T$ and  $\mathbf{0}\in\mathbb{R}^T$. In what follows, we consider a financial policy of \textit{net metering} \cite{nimalsiri2021coordinated}, where customers are compensated for delivering power to the grid and billed for consuming power from the grid at the same rate. The electricity \textit{price profile} is represented by $\bm{\eta} \coloneqq [\eta(1),\dots,\eta(t), \dots, \eta(T)]^\top\in\mathbb{R}^T$, where $\eta(t)$ is the price of electricity (in \$/kWh) at time $t\Delta$. Next we define the \textit{operational cost} (in \$/day) accrued to $\text{EV}_n$ by
\begin{equation}\label{single_user_oper_cost}
   \Omega_n(\bm{x}_n) \ \coloneqq \ \textstyle \sum_{t=1}^T \{\Delta \eta(t) x_n(t) +  \kappa_n x_n(t)^2\},
\end{equation}
where $\kappa_n>0$ (in \$/kW$^{2}$) is a (small) regularization parameter that is introduced to avoid the occurrence of unnecessary charge-discharge cycles, and $\kappa_n x_n(t)^2$ acts as a proxy for the battery degradation cost.

\subsection{Network-aware EV operation}

We consider a three-phase, unbalanced, radial distribution feeder in which the set of nodes is $\mathcal{K}_{0}\coloneqq\{0,\dots,k,\dots,K\}$ (with node $0$ representing the feeder head) and the set of line segments is $\mathfrak{E} \coloneqq \{(k\hat{k})_{k,\hat{k}\in\mathcal{K}_{0}}\}$ $\subseteq\mathcal{K}_{0} \times\mathcal{K}_{0}$ with $|\mathfrak{E}| = K$. Each node and each line can be single-phase, two-phase, or three-phase, with each phase $\phi \in \{a, b, c\}$. The phases along line $(k\hat{k})\in\mathfrak{E}$ is denoted by the set $\Phi^{(k\hat{k})}$. For each node  $k\in\mathcal{K}\coloneqq\mathcal{K}_0\backslash \{0\}=\{1,\dots,K\}$, the line segments on the path from node $0$ to node $k$ is denoted by the set $\mathfrak{E}^k\subseteq\mathfrak{E}$. The set of phases at node $k\in\mathcal{K}_{0}$ is denoted by $\Phi^k$ (e.g., $\Phi^0 = \{a, b, c\}$). We define $\mathbb{K} \coloneqq \cup_{k\in\mathcal{K}} \{(k:\phi)\}_{\phi\in\Phi^k}$ and $|\mathbb{K}| = \varkappa$, where $(k:\phi)\in\mathbb{K}$ represent a \textit{supply point} (phase $\phi$ at node $k$) that is connected to $N^{(k:\phi)}\geq 0$ number of customers (each of whom is associated with a RES in Fig.~\ref{fig-1}), such that $\textstyle\sum_{k\in\mathcal{K}}\sum_{\phi\in\Phi^k} {N^{(k:\phi)}} = N$. Let $\bm{\Upsilon}\in\mathbb{R}^{\varkappa \times N}$, with rows indexed by elements of $\mathbb{K}$ and columns indexed by elements of $\mathcal{N}$, such that $[\bm{\Upsilon}]_{(k:\phi),n}=1$ if customer $n$ is connected to supply point $(k:\phi)\in\mathbb{K}$ and $[\bm{\Upsilon}]_{(k:\phi),n}=0$ otherwise. 

Let ${z}^{k\hat{k},\phi\hat{\phi}} \coloneqq \mathrm{r}^{k\hat{k},\phi\hat{\phi}} + \mathfrak{i} \mathrm{x}^{k\hat{k},\phi\hat{\phi}}$ ($\mathfrak{i}^{2}=-1$), where $\mathrm{r}^{k\hat{k},\phi\hat{\phi}}$ and $\mathrm{x}^{k\hat{k},\phi\hat{\phi}}$ denote the resistance and reactance of line $(k\hat{k})$, with $\phi$ and $\hat{\phi}$ being phases of the line. Then we define $\mathcal{Z}^{k\hat{k},\phi\hat{\phi}}:=\sum_{(\dot{k}\ddot{k})\in \left(\mathfrak{E}^{k}\cap\mathfrak{E}^{\hat{k}}\right)}z^{\dot{k}\ddot{k},\phi\hat{\phi}}$. Let $\bm{R} \in \mathbb{R}^{\varkappa\times \varkappa}$ and $\bm{X} \in \mathbb{R}^{\varkappa\times \varkappa}$ be two square matrices with rows and columns indexed by elements of $\mathbb{K}$, such that $[\bm{R}]_{(k:\phi),(\hat{k}:\hat{\phi)}} \coloneqq  2\mathbb{R}\text{e}\{{(\mathcal{Z}^{k\hat{k},\phi\hat{\phi}})}^*\cdot\omega^{[[\phi]]-[[\hat{\phi}]]}\} \in \mathbb{R}$ and $[\bm{X}]_{(k:\phi),(\hat{k}:\hat{\phi})} \coloneqq -2 \mathbb{I}\text{m}\{{(\mathcal{Z}^{k\hat{k},\phi\hat{\phi}})}^*\cdot\omega^{[[\phi]]-[[\hat{\phi}]]}\} \in \mathbb{R}$ for $k,\hat{k}\in\mathcal{K}$, $\phi\in\Phi^k$, $\hat{\phi}\in\Phi^{\hat{k}}$. Here, $\mathbb{R}\text{e}\{\cdot\}$ and  $\mathbb{I}\text{m}\{\cdot\}$ are the real and imaginary parts, $(\cdot)^*$ is the complex conjugate, $\omega=e^{-\frac{2\pi\mathrm{i}}{3}}$, $[[a]]=0$, $[[b]]=1$, and $[[c]]=2$. 

First, we model temporal voltage evolution across the unbalanced distribution grid. We define $\bm{V}^0 \coloneqq v^{0}\bm{1} \in\mathbb{R}^{\varkappa}$, where $v^0$ is the squared voltage magnitude of each phase at node $0$, which is set to the squared nominal voltage. Let $v^{(k:\phi)}(t)$ be the squared voltage magnitude at supply point $(k:\phi)$ (i.e., phase $\phi\in\Phi^k$ at node $k\in\mathcal{K}$) at time $t\Delta$, and let $\bm{v}^k(t) \coloneqq [\{v^{(k:\phi)}(t)\}_{\phi\in\Phi^k}]^\top\in\mathbb{R}^{|\Phi^k|}$. Then, by concatenating squared voltages across all supply points in $\mathbb{K}$ at time $t\Delta$, we define $\bm{V}(t) \coloneqq [\bm{v}^{1}(t)^\top, \dots,\bm{v}^{t}(t)^\top, \dots,\bm{v}^{K}(t)^\top]^\top \in \mathbb{R}^{\varkappa}$. 

Let $p^{(k:\phi)}(t)$ [kW] be the net real power  (consumption if positive and injection if negative) at supply point $(k:\phi)\in\mathbb{K}$ at time $t\Delta$, with $\widetilde{p}^{(k:\phi)}(t)$ representing the contribution by all non-EV loads and $\widehat{p}^{(k:\phi)}(t)$ representing the contribution by all EVs. Similarly, let $q^{(k:\phi)}(t)$ [kVAR] be the net reactive power (consumption if positive and injection if negative) at supply point $(k:\phi)\in\mathbb{K}$ at time $t\Delta$, where $\widetilde{q}^{(k:\phi)}(t)$ corresponds to the contribution by non-EV loads and $\widehat{q}^{(k:\phi)}(t)$ corresponds to the contribution by EVs. Accordingly, we define 
$\bm{p}^k(t) \coloneqq [\{p^{k: \phi}\}_{\phi\in\Phi^k}(t)]^\top$, $\bm{q}^k(t) \coloneqq [\{q^{k: \phi}\}_{\phi\in\Phi^k}(t)]^\top$, $\bm{\widetilde{p}}^k(t) \coloneqq [\{\widetilde{p}^{k: \phi}\}_{\phi\in\Phi^k}(t)]^\top$, $\bm{\widetilde{q}}^k(t) \coloneqq [\{q^{k: \phi}\}_{\phi\in\Phi^k}(t)]^\top$, $\bm{\widehat{p}}^k(t) \coloneqq [\{\widehat{p}^{k: \phi}\}_{\phi\in\Phi^k}(t)]^\top$ and $\bm{\widehat{q}}^k(t) \coloneqq [\{\widehat{q}^{k: \phi}\}_{\phi\in\Phi^k}(t)]^\top$ such that $\bm{P}(t) \coloneqq [\bm{p}^{1}(t)^\top, \dots, \bm{p}^{K}(t)^\top]^\top\in \mathbb{R}^{\varkappa}$, $\bm{Q}(t) \coloneqq [\bm{q}^{1}(t)^\top, \dots , \bm{q}^{K}(t)^\top]^\top\in \mathbb{R}^{\varkappa}$, $\bm{\widetilde{P}}(t) \coloneqq [\bm{\widetilde{p}}^{1}(t)^\top, \dots, \bm{\widetilde{p}}^{K}(t)^\top]^\top\in \mathbb{R}^{\varkappa}$, $\bm{\widetilde{Q}}(t) \coloneqq [\bm{\widetilde{q}}^{1}(t)^\top, \dots, \bm{\widetilde{q}}^{K}(t)^\top]^\top\in \mathbb{R}^{\varkappa}$, $\bm{\widehat{P}}(t) \coloneqq [\bm{\widehat{p}}^{1}(t)^\top,\dots, \bm{\widehat{p}}^{K}(t)^\top]^\top\in \mathbb{R}^{\varkappa}$ and $\bm{\widehat{Q}}(t) \coloneqq [\bm{\widehat{q}}^{1}(t)^\top,\dots, \bm{\widehat{q}}^{K}(t)^\top]^\top \in \mathbb{R}^{\varkappa}$. According to the LinDistFlow power flow equations in \cite{arnold2016optimal} that are proposed for an unbalanced distribution feeder
\begin{equation}\label{ldf}
    \bm{V}(t) \coloneqq \bm{V}^0 - \bm{R}\bm{P}(t) - \bm{X}\bm{Q}(t).
\end{equation}

Ignoring losses as in \cite{arnold2016optimal} gives $p^{(k:\phi)}(t) = \widetilde{p}^{(k:\phi)}(t) + \widehat{p}^{(k:\phi)}(t)$ and $q^{(k:\phi)}(t) = \widetilde{q}^{(k:\phi)}(t) + \widehat{q}^{(k:\phi)}(t)$ for all $\phi\in\Phi^{k}$ and $k\in\mathcal{K}$, implying $\bm{P}(t) =  \bm{\widetilde{P}}(t) + \bm{\widehat{P}}(t)$ and  $\bm{Q}(t) =  \bm{\widetilde{Q}}(t) + \bm{\widehat{Q}}(t)$ for all $t\in\mathcal{T}$. The squared voltage deviation arising from the aggregate non-EV load (baseline load) at time $t\Delta$ is expressed by $\bm{\widetilde{V}}(t)  \coloneqq \bm{V}^0 - \bm{R}\bm{\widetilde{P}}(t) - \bm{X}\bm{\widetilde{Q}}(t)$. For simplicity, we assume that EVs only charge and discharge real power. Then $\widehat{q}^{(k:\phi)}(t)=0$ for each $(k:\phi)\in \mathbb{K}$, implying $\bm{\widehat{Q}}(t)=\mathbf{0}$ for all $t\in\mathcal{T}$. Let $\bm{\mathcal{X}}(t) \coloneqq [x_1(t), \dots ,  x_n(t) ,\dots, x_N(t)]^\top\in\mathbb{R}^N$ such that $\bm{\widehat{P}}(t) \coloneqq \bm{\Upsilon} \bm{\mathcal{X}}(t)$. Let $\bm{D} \coloneqq  - \bm{R}\bm{\Upsilon} = [\bm{D}_1, \dots, \bm{D}_n ,\dots, \bm{D}_N ]$. Then Eq.~\eqref{ldf} becomes $\bm{V}(t) = \bm{\widetilde{V}}(t) + \bm{D}\bm{\mathcal{X}}(t)$
    
Let us now expand $\bm{V}(t)$ along a time-horizon $[0,T\Delta]$.
\begin{equation}\label{ldf4}
    \bm{V} = \bm{\widetilde{V}} + \textstyle\sum_{n=1}^N \overline{\bm{D}}_n \bm{x}_n,
\end{equation}
where $\bm{V} \coloneqq [\bm{V}(1)^\top,\dots, \bm{V}(T)^\top]^\top\in\mathbb{R}^{\varkappa T}$, $\bm{\widetilde{V}} \coloneqq [\bm{\widetilde{V}}(1)^\top, \dots, \bm{\widetilde{V}}(T)^\top]^\top\in\mathbb{R}^{\varkappa T}$ and $\bm{\overline{D}}_n \coloneqq \oplus_1^T\bm{D}_n \in\mathbb{R}^{\varkappa T\times T}$. 

Next, we model the transformer's thermal dynamics. The transformer that connects the distribution feeder to the
high-voltage transmission network is modeled as a single
thermal mass with following discrete-time temperature dynamics \cite{hermans2012incentive}. 
\begin{equation}\label{temp2}
    \theta(t+1) =  \varrho\theta(t) + \hat{\varrho}i^2(t) + \bar{\varrho}\theta_a(t),
\end{equation}
where aggregated current $i(t) \coloneqq i_d(t) + \textstyle\sum_{n=1}^N i_n(t)$, $\varrho\coloneqq 1-\Delta/(RC)$, $\hat{\varrho} \coloneqq \Delta R_c/C$, $\bar{\varrho} \coloneqq \Delta/RC = 1 - \varrho$. Here, $\theta(t)$ [K] is the transformer core temperature at time $t\Delta$, $i_n(t)$ [A] is the charging current of EV $n$,  $C$ [$\text{JK}^{-1}$] is the transformer heat capacity, $R$ [$\text{KW}^{-1}$] is the heat outflow resistance, and $R_c$ [$\Omega$] is the coil resistance. The net non-EV current $i_d(t)$ [A] and the ambient temperature ${\theta}_a(t)$ [K] at time $t\Delta$ act as exogenous disturbances. As in \cite{hermans2012incentive}, linearization around the equilibrium point $\theta^*$, $i^* \coloneqq \sqrt{\hat{\varrho}^{-1}\bar{\varrho}(\theta^*- \theta_a^*)}$ for $ \theta_a \coloneqq \theta_a^*$ and $i_d \coloneqq 0$ yields the approximate transformer dynamics described by $\delta_{\theta(t+1)} =  \varrho\delta_{\theta(t)} + 2\hat{\varrho}i^*\big(i_d(t) + \textstyle\sum_{n=1}^N \delta_{i_n(t)}\big) + \bar{\varrho}\delta_{\theta_a(t)}$ where $\delta_{\theta(t)} \coloneqq \theta(t) - \theta^*$, $\delta_{i_n(t)} \coloneqq i_n(t) - i^*/N$ and $\delta_{\theta_a(t)} \coloneqq \theta_a(t) - \theta_a^*$. Combining with \eqref{temp2}, we obtain
\begin{equation}\label{temp4}
    \theta(t+1) =  \varrho\theta(t) + \widetilde{\varrho} i(t) + \bar{\varrho}\theta_a(t) + \beta,
\end{equation}
where $\widetilde{\varrho} \coloneqq 2\hat{\varrho}i^*$, $\beta \coloneqq (1-\varrho)\theta^* -\widetilde{\varrho}i^* - \bar{\varrho}\theta_a^*$. Augmenting \eqref{temp4} along a time-horizon $[0, T\Delta]$ yields
\begin{equation}\label{temp5}
    \bm{\theta} = \bm{\varrho} \theta_0 + \bm{\Xi}\bm{i} + \bm{\bar{\varrho}}\bm{\theta_a} + \beta\bm{B},
\end{equation}
where $\theta_0$ is the  transformer temperature at the beginning of the time horizon, $\bm{\theta}\coloneqq[\theta(1), \dots, \theta(T)]^\top\in\mathbb{R}^T$, $\bm{\varrho}\coloneqq[\varrho, \varrho^2, \dots, \varrho^T]^\top\in\mathbb{R}^T$,  $\bm{\theta_a}\coloneqq[\theta_a(0), \dots, \theta_a(T-1)]^\top\in\mathbb{R}^T, \bm{i}\coloneqq[i(0), \dots, i(T-1)]^\top\in\mathbb{R}^T, \bm{B}\coloneqq\beta[1, (1+\varrho), (1+\varrho + \varrho^2), \dots, (1+\varrho+\dots+\varrho^{T-1})]^\top\in\mathbb{R}^T$,

\begin{equation*}
    \bm{\Xi}\coloneqq 
\begin{bmatrix} 
     \widetilde{\varrho}           &   0           & \dots  & 0\\
	 \varrho\widetilde{\varrho}       &   \widetilde{\varrho}     & \dots  & 0\\
      \vdots        &  \vdots       & \ddots & \vdots \\
      \varrho^{T-1}\widetilde{\varrho} & \varrho^{T-2}\widetilde{\varrho} & \dots & \widetilde{\varrho}
	\end{bmatrix}\in\mathbb{R}^{T \times T} ,
\end{equation*}

\begin{equation*}
\bm{\bar{\varrho}}\coloneqq\begin{bmatrix} 
     \bar{\varrho}           &   0           & \dots  & 0\\
	 \varrho\bar{\varrho}       &   \bar{\varrho}     & \dots  & 0\\
      \vdots        &  \vdots       & \ddots & \vdots \\
      \varrho^{T-1}\bar{\varrho} & \varrho^{T-2}\bar{\varrho} & \dots & \bar{\varrho}
	\end{bmatrix}\in\mathbb{R}^{T \times T}.
\end{equation*}


In what follows, we formulate the distribution network constraints using the equations \eqref{ldf4} and \eqref{temp5}. First, the voltage at each supply point $(k:\phi)\in\mathbb{K}$ is constrained to stay within an operational range $[ \underline{v}^{(k:\phi)},\overline{v}^{(k:\phi)}]$. Let  $\bm{\underline{v}} \coloneqq [ [\{\underline{v}^{(1:\phi)}\}_{\phi\in\Phi^1}], \dots,  [\{\underline{v}^{(K:\phi)}\}_{\phi\in\Phi^K}]]^\top\in\mathbb{R}^\varkappa$ and $\bm{\overline{v}} \coloneqq [[\{\overline{v}^{(1:\phi)}\}_{\phi\in\Phi^1}], \dots,  [\{\overline{v}^{(K:\phi)}\}_{\phi\in\Phi^K}]]^\top\in\mathbb{R}^\varkappa$ , such that $\bm{\underline{V}} \coloneqq  [\bm{\underline{v}}^\top, \dots, \bm{\underline{v}}^\top]^\top \in\mathbb{R}^{\varkappa T}$ and $\bm{\overline{V}} \coloneqq [\bm{\overline{v}}^\top, \dots, \bm{\overline{v}}^\top]^\top \in\mathbb{R}^{\varkappa T}$. Further, let $\bm{\mathcal{D}}_n \coloneqq [\overline{\bm{D}}_n^\top \  -\overline{\bm{D}}_n^\top]^\top\in\mathbb{R}^{2\varkappa T\times T}$ and $\bm{\mathfrak{W}} \coloneqq [(\bm{\overline{V}} - \bm{\widetilde{V}})^\top \  (-\bm{\underline{V}} + \bm{\widetilde{V}})^\top]^\top\in\mathbb{R}^{2\varkappa T}$. Next a \textit{voltage constraint} $\bm{\underline{V}} \leq \bm{V} \leq   \bm{\overline{V}}$ is enforced and combined with \eqref{ldf4} to obtain
\begin{equation}\label{voltage_cons_simplified}
    \textstyle \sum_{n=1}^N \bm{\mathcal{D}}_n \bm{x}_n \leq \bm{\mathfrak{W}}.
\end{equation}

Next, to prevent transformer overheating, a \textit{thermal constraint} is imposed as $\bm{\theta} \leq \theta^{max}\bm{1}_T$, where $\theta^{max}$ is an upper bound for the core temperature of the transformer. Then by ~\eqref{temp5}, we have
\begin{equation}\label{temp_cons3}
      \textstyle\sum_{n=1}^N \bm{\Xi} \bm{x}_n  \leq \bm{\mathfrak{J}},
\end{equation}
where $\bm{\mathfrak{J}}\coloneqq v'(\theta^{max}\bm{1}_T - \bm{\varrho} \theta_0 - \bm{\bar{\varrho}}\bm{\theta_a} - \beta^*\bm{B} - \bm{\Xi} \bm{i}_d)\in\mathbb{R}^T$ with $v'$ representing the rms grid voltage.

Combining the voltage and thermal network constraints \eqref{voltage_cons_simplified} and \eqref{temp_cons3} yields 
\begin{equation}\label{combined_cons}
    \textstyle \sum_{n=1}^N \bm{\Gamma}_n \bm{x}_n \leq \bm{w},
\end{equation}
where $\bm{\Gamma}_n \coloneqq [\bm{\Xi}^\top \  \bm{\mathcal{D}}_n^\top]^\top\in\mathbb{R}^{(T+2\varkappa T)\times T}$ and $\bm{w} \coloneqq [\bm{\mathfrak{J}}^\top \  \bm{\mathfrak{W}}^\top]^\top\in\mathbb{R}^{T+2\varkappa T}$. With the above as background, our primary optimization problem $(\mathcal{P}1)$ becomes
\begin{taggedsubequations}{$\mathcal{P}1$}\label{p1}
\begin{align}
\min_{\{\bm{x}_n\}\in\mathbb{R}^T} & \quad \textstyle\sum_{n=1}^N \Omega_n(\bm{x}_n)\label{p1.1}\\ 
\text{s.t.} & \quad \bm{x}_n\in\Psi_n \quad\quad; \ \forall n\in\mathcal{N}, \label{p1.2}\\
 & \quad \textstyle\sum_{n=1}^N \bm{\Gamma}_n \bm{x}_n \leq \bm{w},\label{p1.3}%
\end{align}
\end{taggedsubequations}

\noindent The objective function in \eqref{p1.1} minimizes the operational costs of all customers, subject to the charge-discharge feasibility sets $\Psi_n$ \eqref{p1.2} in \eqref{aggregate_battery_cons} and the coupled network constraints \eqref{p1.3} defined in \eqref{combined_cons}. 

\subsection{An ADMM-based Approach for EV Coordination}

Consider a peer-to-peer communication network represented by an undirected graph $\mathcal{G}:=\{\mathcal{N},\mathcal{E}\}$, where the vertex-set $\mathcal{N}$ is the set of customers (referred to as \textit{agents}), and the edge-set $\mathcal{E}\subseteq\mathcal{N}\times{\mathcal{N}}$ is the set of communication links between agents. An edge $(nm)\in\mathcal{E}$ allows bidirectional communication between agent $n$ and agent $m$. The set of neighboring agents with which agent $n$ can communicate is defined by $\mathcal{N}_n \coloneqq \{m \in \mathcal{N} | \ (nm)\in\mathcal{E}\}$. For agent $n\in\mathcal{N}$, let us introduce a slack variable vector $\bm{s}_n\in\mathbb{R}^{T+2\varkappa T}$ and define $\bm{u}_n \coloneqq [\bm{x}_n^\top \ \ \bm{s}_n^\top]^\top\in\mathbb{R}^{2T+2\varkappa T}$ and $\bm{\xi}_n \coloneqq [\bm{\Gamma}_n \ \  \mathbf{I}]\in\mathbb{R}^{(T+2\varkappa T) \times({2T+2\varkappa T})}$. Define the sets $\mathcal{S}_n \coloneqq \{\bm{s}_n | \bm{s}_n\geq\textbf{0}\}$ and $\mathcal{F}_n \coloneqq \{\bm{u}_n | \ \bm{x}_n\in\Psi_n, \bm{s}_n\in\mathcal{S}_n\}$. Define an indicator function $\mathcal{I}_n(\bm{u}_n)$, where $\mathcal{I}_n(\bm{u}_n) = 0$ if $\bm{u}_n\in\mathcal{F}_n$ and $\mathcal{I}_n(\bm{u}_n) = \infty$ otherwise. Define a local cost function $f_n(\bm{u}_n) \coloneqq \Omega_n(\bm{u}_n) + \mathcal{I}_n(\bm{u}_n)$. Then \eqref{p1} is equivalent to:

\begin{taggedsubequations}{$\mathcal{P}2$}\label{p2}
\begin{align}
\underset{\{\bm{u}_n\}\in\mathbb{R}^{2T+2\varkappa T}}{\mathrm{min}}  \  \textstyle\sum_{n=1}^N f_n(\bm{u}_n)\ \quad \quad \quad \quad \label{p2.1} \\
\mathrm{s.t.} \ \ \textstyle\sum_{n=1}^N \bm{\xi}_n \bm{u}_n = \textstyle\sum_{n=1}^N (\bm{\Gamma}_n \bm{x}_n + \bm{s}_n) = \bm{w}. \label{p2.2}%
\end{align}
\end{taggedsubequations}

We now reformulate \eqref{p2} as a distributed consensus-based optimization problem using the duality and the consensus theory \cite{chang2014multi}. First, for each agent $n\in\mathcal{N}$, we define $\varphi_{n}(\bm{\lambda}) = \underset{\bm{u}_n\in\mathbb{R}^{2T+2\varkappa T}}{\mathrm{max}} \big(-f_n(\bm{u}_n) - \bm{\lambda}^\top\bm{\xi}_n \bm{u}_n \big)$,
where $\bm{\lambda}\in\mathbb{R}^{T+2\varkappa T}$ is the global Lagrange dual variable vector corresponding to \eqref{p2.2}. When $\mathcal{G}$ is connected, \eqref{p2} is equivalent to \eqref{p3}:

\begin{taggedsubequations}{$\mathcal{P}3$}\label{p3}
\begin{align}
\underset{\{\bm{\lambda}_n\}, \{\bm{\beta}_{nm}\}\in\mathbb{R}^{T+2\varkappa T}}{\mathrm{min}} & \textstyle\sum_{n=1}^N \big(\varphi_{n}(\bm{\lambda}_n) + \frac{1}{N}\bm{\lambda}_n^\top\bm{w}\big)\label{p3.1}\\
\mathrm{s.t.} \quad & \bm{\lambda}_n = \bm{\beta}_{nm} ; \forall m\in\mathcal{N}_n,  n\in\mathcal{N},\label{p3.2}\\
\quad & \bm{\lambda}_m = \bm{\beta}_{nm}; \forall m\in\mathcal{N}_n,  n\in\mathcal{N},\label{p3.3}%
\end{align}
\end{taggedsubequations}

\noindent where $\bm{\lambda}_{n}\in\mathbb{R}^{T+2\varkappa T}$ is the $n^{\text{th}}$ customer's local copy of global dual variable $\bm{\lambda}$ and $\{\bm{\beta}_{nm}\}_{n\in\mathcal{N}, m\in \mathcal{N}_n}\in\mathbb{R}^{T+2\varkappa T}$ are auxiliary consensus variables. Let $\{\bm{\nu}_n\}_{n\in\mathcal{N}}\in\mathbb{R}^{T+2\varkappa T}$ be the set of local dual variable vectors corresponding to auxiliary consensus constraints \eqref{p3.2}-\eqref{p3.3}. The iterative ADMM updates $\bm{\nu}_n$ at each iteration $\tau > 0$ as follows.
\begin{equation}\label{up1}
    \bm{\nu}_n^{[\tau]} = \bm{\nu}_n^{[\tau-1]} + \rho \textstyle\sum_{m\in\mathcal{N}_n} (\bm{\lambda}_n^{[\tau-1]} - \bm{\lambda}_m^{[\tau-1]}),
\end{equation}
where $\rho>0$ is a penalty parameter. Next, $\bm{u}_n$ and $\bm{\lambda}_n$ are updated as below.




\begin{multline}\label{up3}
    \bm{u}_n^{[\tau]} = \text{arg} \min_{\bm{u}_n\in\Xi_n} \Bigl\{  f_n(\bm{u}_n)  + \textstyle\frac{\rho}{4|\mathcal{N}_n|} \Big\lVert \frac{1}{\rho} \Big(\bm{\xi}_n\bm{u}_n - \frac{1}{N}\bm{w} \Big)\\
    -\frac{1}{\rho} \bm{\nu}_n^{[\tau]} + \sum_{m\in\mathcal{N}_n} \Big( \bm{\lambda}_n^{[\tau-1]} + \bm{\lambda}_m^{[\tau-1]} \Big)\Big\rVert^2_2 \Bigl\},
\end{multline}
  
\begin{multline}\label{up4}
   \noindent\bm{\lambda}_n^{[\tau]} = \frac{1}{2|\mathcal{N}_n|} \Bigl\{ \sum_{m\in\mathcal{N}_n} (\bm{\lambda}_n^{[\tau-1]} + \bm{\lambda}_m^{[\tau-1]})  - \frac{\bm{\nu}_n^{[\tau]}}{\rho}  + \\ \frac{\bm{\xi}_n\bm{u}_n^{[\tau]}}{\rho} - \textstyle\frac{\bm{w}}{N\rho} \Bigl\}.
\end{multline}

\begin{algorithm}[t]
\caption{\hbox{}}
\nonl \textbf{Given}: Initial local variables $\bm{\lambda}_n^{[0]} = \bm{0}$, $\bm{\nu}_n^{[0]} = \bm{0}$ $\forall n\in\mathcal{N}$, iteration index $\tau=1$\\
    \Repeat 
    {    
    \ForEach {agent $n\in\mathcal{N}_a^{[\tau]}$ (in parallel)}{\label{rs}
            \ForEach{neighbor $m\in\mathcal{N}_n$}{
            \uIf{$m\in\{m | (nm)\in\mathcal{E}_a^{[\tau]}\}$}
            {Receive  $\bm{\lambda}_m^{[\tau-1]}$ through link $(nm)$.}
            \uElse{Consider $\bm{\lambda}_m^{[\tau-1]} = \bm{\lambda}_m^{[\tau-2]}$.}
            }
            Update  $\bm{\nu}_n^{[\tau]}$ as per \eqref{up1}.\label{up_nu}\\
            With $\bm{\nu}_n^{[\tau]}$ from line~\ref{up_nu}, update $\bm{u}_n^{[\tau]}$ as per \eqref{up3}.\label{up_u}\\
            With  $\bm{\nu}_n^{[\tau]}$ from line~\ref{up_nu} and $\bm{u}_n^{[\tau]}$ from  line~\ref{up_u}, update $\bm{\lambda}_n^{[\tau]}$ as per \eqref{up4}.\label{up_lambda}\\
            Broadcast $\bm{\lambda}_n^{[\tau]}$ to all neighbors in $\mathcal{N}_n$.\label{end_routine}\\
    }
    \ForEach {agent $n\notin\mathcal{N}_a^{[\tau]}$ (in parallel)}{
            Set  $\bm{\nu}_n^{[\tau]} = \bm{\nu}_n^{[\tau-1]}$, $\bm{u}_n^{[\tau]} = \bm{u}_n^{[\tau-1]}$, $\bm{\lambda}_n^{[\tau]}=\bm{\lambda}_n^{[\tau-1]}$.
    }
        $\tau \coloneqq \tau + 1$.\label{re}\\
    }
    \textbf{until} a stopping criterion is satisfied\;
\end{algorithm}

Next, we allow random updates in the iterative ADMM routine. That is, the agents may stay connected (active) or disconnected (inactive) randomly and the communication between agents may fail probabilistically across different iterations. At each iteration $\tau$, the probabilities of agent $n$ being active is $\hat{\alpha_n}\in(0,1]$ and being inactive is $(1-\hat{\alpha}_n)$. Moreover, for each link $(nm)\in\mathcal{E}$, the probability that the message exchange between agent $n$ and agent $m$ fails (for example, due to packet losses) is $\bar{\alpha}^{(nm)}\in(0,1]$. Therefore, the probability that agents $n$ and $m$ are both active and have successful communication is given by $\hat{\alpha_n}\hat{\alpha_m}(1-\bar{\alpha}^{(nm)})$ --- in which case the link $(nm)\in\mathcal{E}$ is considered active. Accordingly, at each iteration $\tau$, the set of active agents is denoted by $\mathcal{N}_a^{[\tau]}\subseteq\mathcal{N}$ and the set of active links is denoted by $\mathcal{E}_a^{[\tau]}\subseteq\{(nm)\in\mathcal{E}|n,m\in\mathcal{N}_a^{[\tau]}\}$. 

Algorithm 1 summarizes the proposed algorithm under random communication failures. As the \textit{iterative ADMM routine} in lines~\ref{rs}-\ref{re} progresses, the sequence of dual variables $\{\bm{\lambda}_n^{[\tau]}\}_{n\in\mathcal{N}}$  converges to an optimal solution $\bm{\lambda}^*$ and the sequence of primal variables, $\{\bm{u}_n^{[\tau]}\}_{n\in\mathcal{N}}$, converges to an optimal solution $\{\bm{u}_n^*\}_{n\in\mathcal{N}}$, following the Theorem 2 of \cite{chang2014multi}. More importantly, Algorithm~1 is guaranteed to converge to an optimal solution even under nonideal network conditions, following Theorem 1 of \cite{chang2015randomized}.

\section{Numerical Simulations}

The following numerical simulations are conducted using the IEEE 13-node test feeder, a representative unbalanced distribution circuit. The test feeder is populated with $600$ EVs connected by a random graph $\mathcal{G}$ that is generated using the same technique as in \cite{chang2015randomized}. The algorithm is run over a 24-hour time horizon with 30-mins time steps ($\Delta = 0.5$). A voltage threshold of $\pm 4.6$\% about the nominal voltage $1~\text{p.u.}$ is considered. An upper bound of $\theta^{max}+\theta^* = 393 K$ is considered for the transformer core temperature, with the remaining transformer and thermal parameters as the same as in \cite{hermans2012incentive}. All of the data corresponding to EV attributes, residential load, PV generation, and electricity pricing are consistent with the simulation setting described in \cite{nimalsiri2021distributed}.

\begin{figure}[t]
    \centering
    \includegraphics[width=\linewidth]{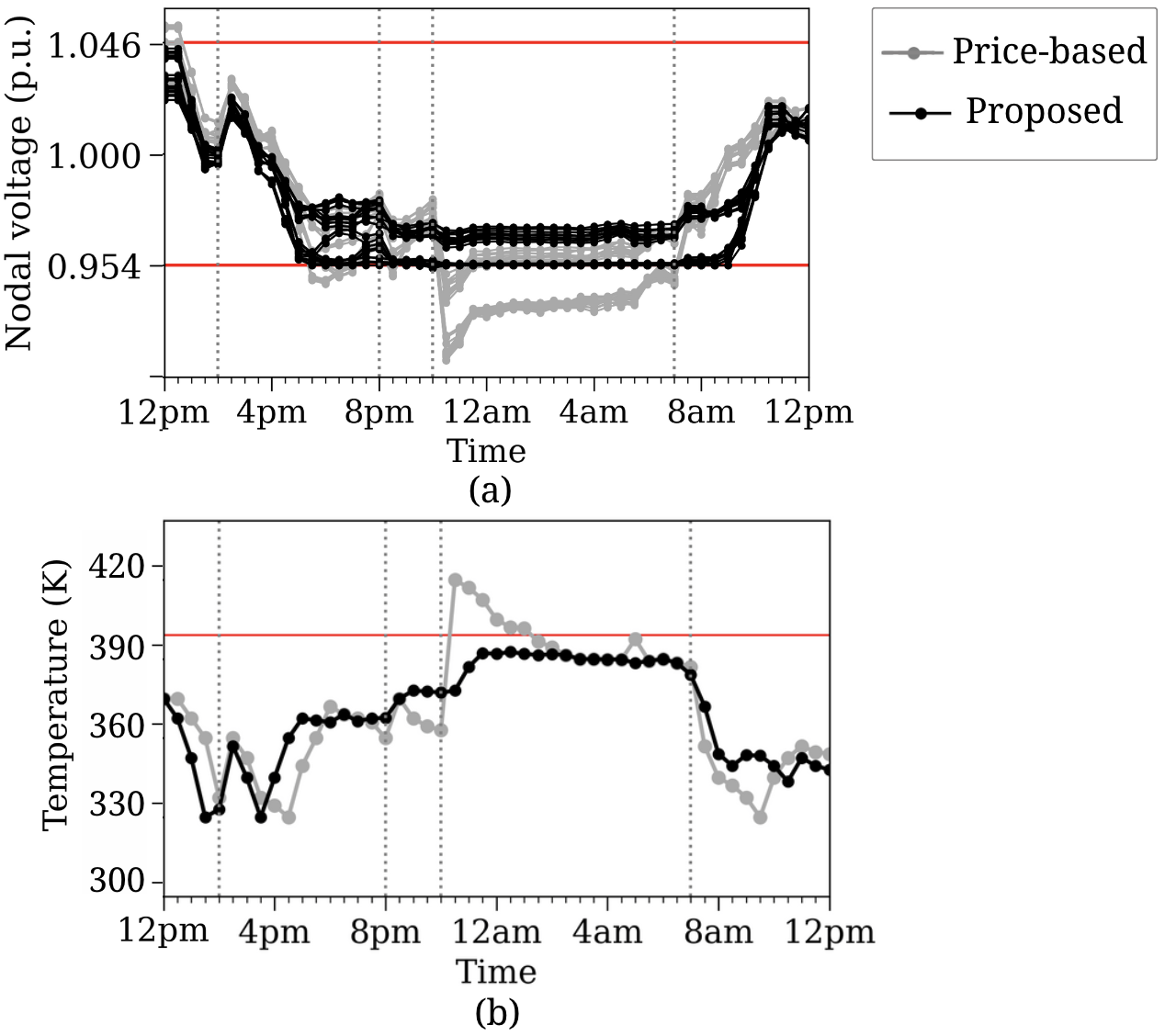}
    \caption{(a) Temporal evolution of voltages at supply points in $\mathbb{K}$; (b) temporal evolution of the transformer core temperature, for Case 1 (price-based) and Case 2 (proposed). Vertical lines indicate ToU pricing, where the off-peak is 10 pm – 7 am, the shoulder-peak is 7 am – 2 pm and 8 pm – 10 pm, and the peak is 2 pm – 8 pm. Red lines mark the operational limits.}
    \label{fig-2}
\end{figure}

In what follows, we consider two cases:\textbf{ Case 1 (Price-based)}: EVs minimize their operational costs without conforming to network constraints (i.e., minimize \eqref{p1.1} subject to \eqref{p1.2}); \textbf{Case 2 (Proposed)}: EVs follow Algorithm~1 to minimize their operational costs in compliance with network constraints (i.e., minimize \eqref{p1.1} subject to \eqref{p1.2}-\eqref{p1.3}).

Fig.~\ref{fig-2}(a) shows the resultant voltage profile at each supply point $(k:\phi)\in\mathbb{K}$, defined by ${\bm{v}^{(k:\phi)}}^{1/2} \coloneqq [{v^{(k:\phi)}(1)}^{1/2}, \dots, {v^{(k:\phi)}(T)^{1/2}}]^\top$, and Fig.~\ref{fig-2}b) shows the transformer core temperature evolution, defined by $\bm{\theta} \coloneqq [\theta(1), \dots, \theta(T)]^\top$, with grey dotted lines corresponding to Case~1 and black dotted lines corresponding to Case~2. Clearly, the voltage profiles resulting from Case~1 drop to voltages below the minimum voltage limit within the off-peak and shoulder-peak periods (as a result of excessive charging load in response to low electricity prices) and rise to voltages above the maximum voltage limit within 12~pm -- 12.30~pm (as a result of high PV generation). Moreover, in Fig.~\ref{fig-2}(b), the transformer core temperature exceeds the preset upper bound within 10.30~pm -- 1.30~pm. By contrast, the solution obtained for Case~2 satisfies the safe voltage and thermal operational limits, while also managing to satisfy all EV charging requirements ahead of their specified departure times.

\begin{figure}[t]
    \centering
    \includegraphics[width=\linewidth]{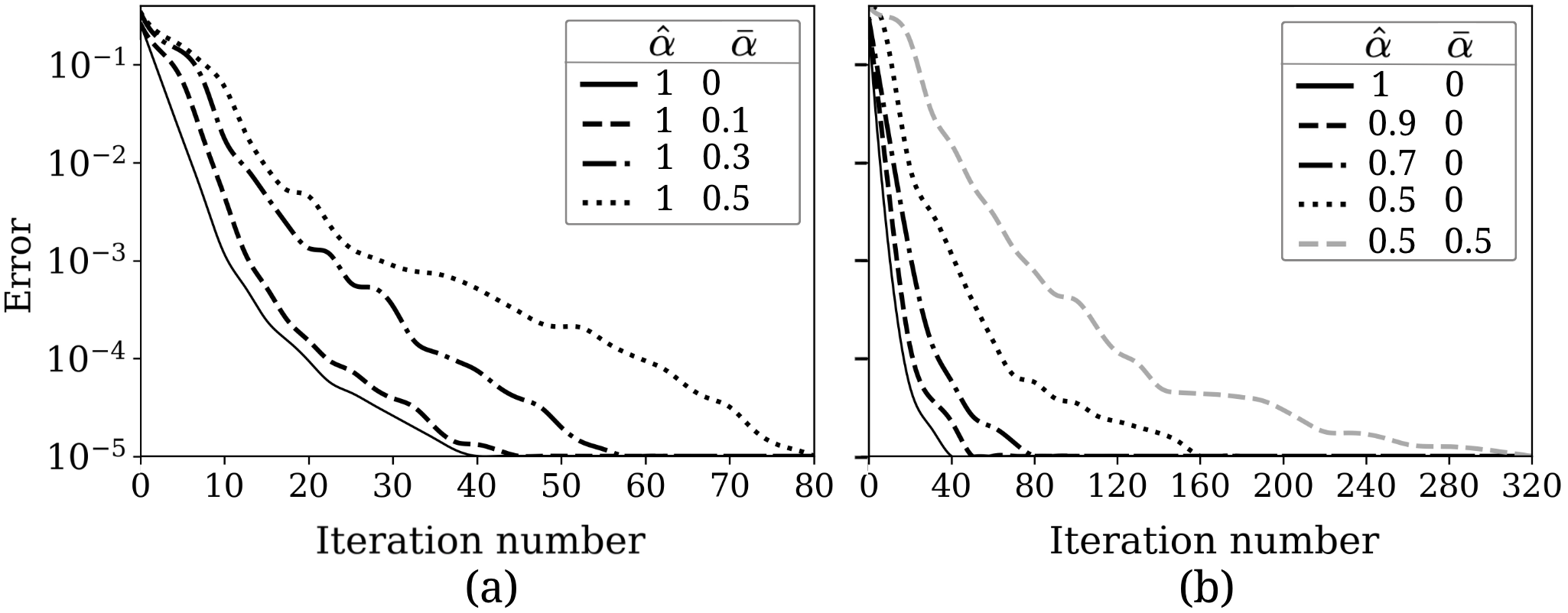}
    \caption{Convergence of Algorithm~1 in the presence of communication failures. Here, $\text{Error} \coloneqq \textstyle\frac{(obj^{[\tau]} - obj^*)}{obj^*}$, where $obj^*$ is the optimal objective value of \eqref{p1.1} and $obj^{[\tau]}$ is the objective value of \eqref{p1.1} at iteration $\tau$.}
    \label{fig-3}
\end{figure}

Next, we examine the behavior of the proposed algorithm against communication network failures. For simplicity, the active probabilities of all agents are considered the same $\hat{\alpha} \triangleq \hat{\alpha}_1 = \dots = \hat{\alpha}_1$ and the link failure probabilities of all edges are considered the same $\bar{\alpha} \triangleq \bar{\alpha}^{(nm)} \forall (nm)\in\mathcal{E}$. Fig.~\ref{fig-3} demonstrates the convergence of Algorithm~1 along the iterative ADMM routine. Specifically, Fig.~\ref{fig-3}(a) considers $\hat{\alpha} = 1$ and shows the convergence for various values of $\bar{\alpha}$, where $\bar{\alpha}=0$ corresponds to the \textit{ideal case} that is free of communication failures. We observe that Algorithm~1 converges consistently through iterations, although communication link failures slow down the convergence rate. In particular, we observe that the number of iterations required to reach an accuracy of $10^{-5}$ is roughly doubled with $\bar{\alpha} = 0.5$ compared to that with $\bar{\alpha} = 0$ --- which is acceptable as with $\bar{\alpha} = 0.5$, the probability of successful message exchange between two agents is reduced by half. Fig.~\ref{fig-3}(b) demonstrates the convergence of Algorithm~1 for various values of $\hat{\alpha}$ with $\bar{\alpha} = 0$. We observe that Algorithm~1 still converges consistently, although the convergence rate is slowed due to random disconnection of agents. Furthermore, we observe that with $\hat{\alpha}=0.5$, the number of iterations required to converge is around four times that for $\hat{\alpha}=1$ --- because the probability of successful message exchange between two agents that are active simultaneously is decreased to one-fourth. Fig.~\ref{fig-3}(b) also depicts the convergence of Algorithm~1 in the events of both inactive agents and communication link failures with $\hat{\alpha}=0.5$ and $\bar{\alpha} = 0.5$, in which case Algorithm~1 converges at a much slower rate ($8$ times slower compared to the ideal case). Importantly, the simulation results in Fig.~\ref{fig-3} imply that: (1) Algorithm~1 performs sufficiently well for nonideal and asynchronous communication networks where agents need not be synchronized with each other; (2) the consensus amongst agents is achieved in spite of random communication uncertainties, verifying the communication-failure tolerant nature of Algorithm~1.

\section{Conclusion}

We have presented a distributed control algorithm to coordinate EV charging and discharging in compliance with distribution network operating limits, while also taking into account communication uncertainties inherent in distributed communication-based approaches. The proposed algorithm satisfied EV charge requirements with minimal operational costs while maintaining the network voltage and thermal limits within statutory limits. Numerical simulations have further verified the practical convergence behavior of the algorithm in the presence of communication failures. The proposed algorithm can be readily extended to EV-based inverter reactive power control to further improve voltage regulation.

\bibliographystyle{IEEEtran}


\end{document}